\newif \ifmac
   \macfalse

\newif \iftwoe 
\ifmac
     \twoefalse
\else
     \twoetrue
\fi

\newif \ifgd
   \gdfalse

\newif\ifdraft
   \draftfalse

\newcommand{\cn}{\sf} 
\newcommand{\kw}{\bf} 

\ifdraft
   
\else
    
\fi

\ifdraft
   \newcommand{\note}[1]{\marginpar{
   \renewcommand{\baselinestretch}{0.8} \footnotesize\sf #1}}
\else
   \newcommand{\note}[1]{}
\fi

\iftwoe
  \documentclass[letterpaper,10pt]{article}
  \usepackage{fullpage}
  \usepackage{times} 
  \usepackage{epic}
  \usepackage{eepic}
  \usepackage{list}  
  \usepackage{wrapfig}
  \usepackage{moreverb}
  \usepackage{alltt}
  \usepackage[nolineno,norules]{lgrind}
\else
  \ifgd
    \documentstyle[psfig]{cmpllncs}
  \else
    \documentstyle[11pt,psfig]{article}
  \fi
\fi

\title{Object-Oriented Design of Graph Oriented Data Structures%
\thanks{Research supported in part by the ESPRIT LTR Project no. 20244 - ALCOM-IT. }
~\\ \vskip 3pt \large                (Extended Abstract)
}

\author{
\addtocounter{footnote}{+1}
\smallskip     {\em Maurizio Pizzonia}%
   \thanks{Dipartimento di Informatica e Automazione,
           Universit\`a di Roma Tre, via della Vasca Navale 79, 00146 Roma, Italy.}\\
   \small \tt pizzonia@dia.uniroma3.it
\and
\smallskip     {\em Giuseppe Di Battista}\footnotemark[3]\\
   \small \tt gdb@dia.uniroma3.it
}
\date{}

\begin{document}

\ifdraft
\tableofcontents
\newpage
\fi

\maketitle

\begin{abstract}

\note{\bf Revise the abstract!} 

Applied research in graph algorithms and combinatorial structures
needs comprehensive and versatile software libraries. However, the design and
the implementation of flexible libraries are challenging activities.
Among the other problems involved in such a difficult field, a very special
role is played by graph classification issues.

We propose new techniques devised to help the designer and
the programmer in the development activities. Such techniques are especially
suited for dealing with graph classification problems and rely on an
extension of the usual object-oriented paradigm.
In order to support the usage of our approach, we devised an extension of the 
C++ programming language and implemented the corresponding pre-compiler.
\end{abstract}


\section{Introduction}
  \note{\bf Revise the intro!} 

Libraries dealing with combinatorial structures like graphs are quite
often large systems because of the great variety of known algorithms
and because of the possibility of graphs to be classified in a plethora of ways.
Also, most of the existing algorithms require a large amount of
software to be implemented. To give an example, the
GDToolkit~\cite{gdtoolkit_man} system consists of more than $40,000$
lines of code. A limited list of object-oriented systems somehow
related to graphs includes: 
ALF~\cite{IEEETSE::BertolazziBL1995},
ffgraph~\cite{f-tfl-95},
JDSL~\cite{Tamassia97},
Leda~\cite{mn-lpcgc-95}, and
LINK~\cite{bdgss-gdml-98}.

Modern large software systems are usually built using object-oriented
technologies and methodologies. The main motivations of this choice are:
development and maintenance time reduction, reusability of components,
and extensibility of the system. A key issue here is to increase the
abstraction level of the code by modeling the reality of interest
with high-level representation structures.

The above mentioned goals are especially difficult to meet in the
application domain that we are considering. Our opinion is that one 
of the main reasons for this difficulty is in the lack, in currently used
object oriented methods and languages, of representation
primitives suitable for representing complex combinatorial structures.
The experience of designing a large library of graph algorithms offers a clear 
perception of this type of modeling problems, that are especially related to 
the classification of graphs. 
We give an overview and an analysis of such problems in
Section~\ref{classification-problems}.

The main contributions of this paper can be summarized as follows:

\begin{itemize}

\item We propose an extension of the object-oriented paradigm,
specifically tailored to tackle problems arising in the classification
of graphs (Section~\ref{se-eco-paradigm}). Such an extension (called {\em ECO})
consists of
new representation primitives.

\item We show how to use ECO to build large graph libraries. This is done by
suggesting
several design schemas. A project experience conducted with
such design schemas has put in evidence an improved flexibility of the
system and a decrease of the development time. The saving of time was
mainly due to the high abstraction level of the produced
code (Section~\ref{se-usage}).

\item We describe an extension of the C++ programming language 
that embodies the ECO primitives. Further,
we present a pre-compiler that we have implemented for ECO, which gives
an easy way to use the concepts of the new paradigm (Section~\ref{se-supporting}). 

\end{itemize}


\section{Classification Problems for Graphs}
\label{classification-problems}

Classification is a key part in software libraries that manage combinatorial
structures like graphs. In fact, each graph algorithm is suited to
work on a specified class of graphs for which given properties hold.
However, notwithstanding its importance, the actual role of graph
classification is usually reduced by the flexibility and the extensibility
problems that a large hierarchy of graph classes introduces
(see, e.g.~\cite{IEEETSE::BertolazziBL1995}).

In the following we briefly analyze the most, in our opinion, 
relevant aspects of the interplay between graph 
classification problems and the object-oriented paradigm.

\subsection{Inheritance and Subclasses}

Various authors point out that saying that $B$ is a
subclass\footnote{We do not distinguish between classes and types (see
\cite{IB-B961146}) because it is not relevant to our
discussion. However, what we say can be easily extended
to handle such distinction.} of $A$ can have several meanings (for
example see~\cite{Rumbaugh91,Wegner87}).

We refer to {\em extension} subclassing if new information is added to $B$ and each
instance of
$A$ can become an instance of $B$ if suitably equipped with new information 
(e.g.\ in this sense Labeled Graph is a subclass of Graph). In the
following we call {\em extension} the ``added part'' of $B$  
and {\em support} the ``inherited part''.   This is very different
from the case where no new information is added to $B$ and there are
instances of $A$ that are not instances of $B$, i.e.\ $B$ is a strict
subclass of $A$ in the mathematical sense (e.g.\ Connected Graph is a
subclass of Graph). In this case we talk about {\em restriction}
subclassing.

The inheritance mechanisms of most object-oriented languages are well
suited in modeling extension subclassing, but fail in modeling
restriction subclassing.  Namely, the restriction subclassing
infringes the soundness of polymorphism because not all values that
are legal for objects that belong to $A$ are also legal for objects
that belong to $B$. I.e.~$B$ violates the \emph{substitution principle}
(first stated in~\cite{Liskov88}).  Hence, the programmer must either
exploit the exception mechanism when the invariant of $B$ is violated,
or, even worse, must rely on the user to behave correctly (such
policies are used in libraries like Leda~\cite{mn-lpcgc-95},
JDSL~\cite{Tamassia97}, LINK~\cite{bdgss-gdml-98}, and
GDToolkit~\cite{gdtoolkit_man}).  Such sort of situations induce the
programmer to modify $A$ in order to consider particular situations
that are actually related to $B$, thus violating another well known
principle, the \emph{open-closed principle}: ``every module must be
closed to modification and open to extensions'' (first stated
in~\cite{Meyer88}).

Consequently, restriction subclassing cannot be modelled in any elegant way using
standard
inheritance mechanisms, unless we make use of the functional-style design\footnote{A
functional-style design avoids any changes to existing objects; methods that change
the
graph actually produce a new copy of the object. The class of the new copy can also
be
different from the starting class.}, which is known to be rather inefficient. 
\note{dai uno sguardo alla nota}


\subsection{Crossed Classifications}

Complex combinatorial structures like graphs can be classified in a
large number of ways. If more than one independent
classification exist, {\em crossed} classes must be created to give to
the system complete classification capabilities. For example,
Fig.~\ref{fi-cross} shows how combining connectivity, planarity, and
orientation may give raise to several crossed classes, even if none of
such classes adds new features to the inherited ones.  Of
course, the number of crossed classes increases exponentially with the
classification coordinates.

This problem is especially relevant because usual object-oriented
paradigms do not allow objects to belong to more than one class. The
only exception is the usage of multiple inheritance that, however, is
not helpful in decreasing the number of subclasses.

Several authors address this problem. For example, \cite{Rumbaugh91} uses a
``delegation technique'', while~\cite{IB-D963068} uses the ``decorator
pattern''. However, the proposed solutions always show an
overwhelming burden for the programmer and/or a quite tricky game of
conventions and limitations that cannot be expressed in a programming
language.

\ifmac
\else
   \begin{figure}
       \centering{\centering{\setlength{\unitlength}{0.00056868in}
\begingroup\makeatletter\ifx\SetFigFont\undefined%
\gdef\SetFigFont#1#2#3#4#5{%
  \reset@font\fontsize{#1}{#2pt}%
  \fontfamily{#3}\fontseries{#4}\fontshape{#5}%
  \selectfont}%
\fi\endgroup%
{\renewcommand{\dashlinestretch}{30}
\begin{picture}(5019,2649)(0,-10)
\path(867,1767)(2217,2307)
\blacken\path(2116.725,2234.579)(2217.000,2307.000)(2094.441,2290.287)(2116.725,2234.579)
\path(867,957)(867,1497)
\blacken\path(897.000,1377.000)(867.000,1497.000)(837.000,1377.000)(897.000,1377.000)
\path(957,957)(2172,1497)
\blacken\path(2074.527,1420.849)(2172.000,1497.000)(2050.158,1475.678)(2074.527,1420.849)
\path(2217,957)(1317,1497)
\blacken\path(1435.334,1460.985)(1317.000,1497.000)(1404.464,1409.536)(1435.334,1460.985)
\path(2802,957)(3792,1497)
\blacken\path(3701.018,1413.201)(3792.000,1497.000)(3672.287,1465.875)(3701.018,1413.201)
\path(4017,957)(2847,1497)
\blacken\path(2968.527,1473.952)(2847.000,1497.000)(2943.383,1419.474)(2968.527,1473.952)
\path(4107,957)(4107,1497)
\blacken\path(4137.000,1377.000)(4107.000,1497.000)(4077.000,1377.000)(4137.000,1377.000)
\path(1947,282)(912,687)
\blacken\path(1034.681,671.209)(912.000,687.000)(1012.817,615.335)(1034.681,671.209)
\path(3072,282)(4107,687)
\blacken\path(4006.183,615.335)(4107.000,687.000)(3984.319,671.209)(4006.183,615.335)
\path(3882,1767)(2802,2307)
\blacken\path(2922.748,2280.167)(2802.000,2307.000)(2895.915,2226.502)(2922.748,2280.167)
\path(2487,282)(2487,687)
\blacken\path(2517.000,567.000)(2487.000,687.000)(2457.000,567.000)(2517.000,567.000)
\path(2487,1767)(2487,2307)
\blacken\path(2517.000,2187.000)(2487.000,2307.000)(2457.000,2187.000)(2517.000,2187.000)
\path(1407,282)(3747,282)(3747,12)
	(1407,12)(1407,282)
\path(1767,1767)(3297,1767)(3297,1497)
	(1767,1497)(1767,1767)
\path(3342,957)(5007,957)(5007,687)
	(3342,687)(3342,957)
\path(1902,2622)(3117,2622)(3117,2307)
	(1902,2307)(1902,2622)
\path(3432,1767)(4737,1767)(4737,1497)
	(3432,1497)(3432,1767)
\path(1812,957)(3252,957)(3252,687)
	(1812,687)(1812,957)
\path(12,957)(1677,957)(1677,687)
	(12,687)(12,957)
\path(192,1767)(1632,1767)(1632,1497)
	(192,1497)(192,1767)
\put(3612,1587){\makebox(0,0)[lb]{\smash{{{\SetFigFont{8}{9.6}{\rmdefault}{\mddefault}{\updefault}PlanarGraph}}}}}
\put(2262,2397){\makebox(0,0)[lb]{\smash{{{\SetFigFont{8}{9.6}{\rmdefault}{\mddefault}{\updefault}Graph}}}}}
\put(282,1587){\makebox(0,0)[lb]{\smash{{{\SetFigFont{8}{9.6}{\rmdefault}{\mddefault}{\updefault}DirectedGraph}}}}}
\put(102,777){\makebox(0,0)[lb]{\smash{{{\SetFigFont{8}{9.6}{\rmdefault}{\mddefault}{\updefault}DirectedConnected}}}}}
\put(1857,1587){\makebox(0,0)[lb]{\smash{{{\SetFigFont{8}{9.6}{\rmdefault}{\mddefault}{\updefault}ConnectedGraph}}}}}
\put(3477,777){\makebox(0,0)[lb]{\smash{{{\SetFigFont{8}{9.6}{\rmdefault}{\mddefault}{\updefault}ConnectedPlanar}}}}}
\put(1542,102){\makebox(0,0)[lb]{\smash{{{\SetFigFont{8}{9.6}{\rmdefault}{\mddefault}{\updefault}DirectedConnectedPlanar}}}}}
\put(1902,777){\makebox(0,0)[lb]{\smash{{{\SetFigFont{8}{9.6}{\rmdefault}{\mddefault}{\updefault}DirectedPlanar}}}}}
\end{picture}
}}} 
    
       \caption{A hierarchy affected by the crossed classification problem.}
       \label{fi-cross}
   \end{figure}
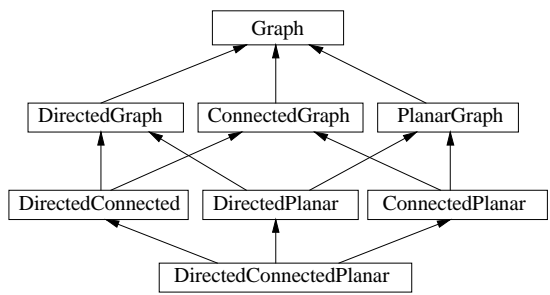
\fi

\subsection{Multiple Decorations}
\label{se-multiple-decoration}

There are two types of subclasses defined with the extension
subclassing.

\begin{enumerate}

\item\label{ca-univocally} The added information is (if needed) univocally
   determined by the support. For example, if the superclass is a graph,
   then the subclass might be a graph equipped with the set of the connected
   components. Given a graph, such information is univocally determined.

\item\label{ca-notunivocally} The added information is not univocally
   determined by the support. For example, an $st$-numbering of a graph is
   (at least partially) arbitrary.

\end{enumerate}

In Case~\ref{ca-univocally}, it is reasonable to exploit the inheritance 
for adding the information about the set of the connected components. 

Case~\ref{ca-notunivocally} is simple if we do not need to add more
information (e.g.\ other $st$-numberings or orientations). If this is
true, then the usage of the inheritance remains a reasonable choice.
However, if we require the support to be extended with other
information of the same type, then it is not possible to use the
inheritance.  In other words, referring to the example, the
inheritance forces to specify just one $st$-numbering, while some
applications may require to $st$-number the vertices of the graph in
several ways at the same time.

A possible solution is to instantiate one new object for each new
information (e.g.\ for each $st$-numbering) which contains the new
information.  It is also needed to state links between elements of the
first object (e.g.\ the vertices) and the new objects (e.g.\ the
numerical labels) by means of suitable structures (e.g.\
hash-tables). Even though many libraries provide mechanisms to state
such links, it may become difficult (or even practically impossible) to maintain the
consistency among the objects when the graph changes.

\subsection{Promotion Efficiency}

During the execution of an algorithm a graph may change its
properties.  This has the effect of virtually moving the graph to
another class, where new methods become meaningful. Further, even if the
properties of a graph do not change, it is sometimes useful, for efficency
reasons, to dynamically equip the graph with new capabilities when they are 
needed.

In general, when an object is {\em promoted} ({\em demoted}) from a
class into its child (parent), it requires a complete copy of the
support. For example, if the graph is recognized to be planar, and we decide to
promote it into the suitable class to take advantage of the new
properties, a complete copy of the graph (with its ``heavy''
implementation structures) must be done.  If the internal
implementation is the same (e.g.\ linked), a large amount of time is wasted to
make the copy.  This happens because the membership relation is
usually not dynamic in the available programming languages.  The class
of an object is statically defined at the moment of its creation.

This problem can be addressed by means of the ``bridge
pattern''~\cite{IB-D963068}. This technique keeps separated the
interface and the implementation structures using two distinct objects
linked by a pointer. In this way, it is very efficient, although not
very elegant, ``to steal'' the implementation from a graph to create
another graph that has an extended interface.  This technique is used
in GDToolkit~\cite{gdtoolkit_man}.


\section{New Representation Primitives for Graphs Classification}
\label{se-eco-paradigm}

The {\em ECO} (Extender and Classer Oriented) paradigm is an extension
of the usual object-oriented design paradigm.  This means that ECO
imports the set of representation primitives usually defined in
object-oriented design methodologies and enlarges such a set with new
concepts.  The new concepts are expecially targeted to address the
problems illustrated in Section~\ref{classification-problems}.

\subsection{Extenders and E-Methods}\label{sec-extender}

The {\em extender} is the main new concept introduced by ECO.
Extenders are classes that have the following additional features.  An
extender is associated to a {\em support-class}. An instance of an
extender, called {\em extension-object}, is created over a {\em
support-object} belonging to the support-class.  It must be created
after the support-object and it must be destroyed before the
support-object. Thus, we can talk about the current extension-objects
of a given support-object. Given a support-class $C$ and an extender $E$ for
$C$, it is possible for an instance of $C$ to be support for more than
one instance of $E$. To give an example, a class  {\cn Graph} can be
support-class for the extenders {\cn Orientation} and {\cn Embedding},
so, more than one orientation and/or embedding may be instantiated over
a given graph.

When an event occurs, then the support-object can notify it to its
current extension-objects. The notification is done by using specific
methods, called {\em E-methods}. An E-method signature is defined in a
support-class and the behavior is specified in its extenders. 
An E-method equips an extension-object with two capabilities. An
extension-object

\begin{itemize}

\item can change its state when the state of the support-object changes,

\item can add constraints to the possible changes of the state of
the support-object (using exception mechanisms).

\end{itemize}

E-methods are invoked only in the methods of the support-class.  The
invocation of an E-method triggers the execution of its behavior for
each current extension-object (see
Fig.~\ref{fi-extender-simple-example}). The behavior of the E-methods
is allowed to modify only the state of the extension-object, which it
is invoked for. The executions sequence is not specified. Intuitively,
they can be considered parallel executions, while the support-object
waits for the termination of all of them.

\ifmac
\else
    \begin{figure}[t]
    \centering{\centering{\setlength{\unitlength}{0.00061242in}
\begingroup\makeatletter\ifx\SetFigFont\undefined%
\gdef\SetFigFont#1#2#3#4#5{%
  \reset@font\fontsize{#1}{#2pt}%
  \fontfamily{#3}\fontseries{#4}\fontshape{#5}%
  \selectfont}%
\fi\endgroup%
{\renewcommand{\dashlinestretch}{30}
\begin{picture}(7044,2829)(0,-10)
\put(2817,297){\arc{210}{1.5708}{3.1416}}
\put(2817,402){\arc{210}{3.1416}{4.7124}}
\put(3102,402){\arc{210}{4.7124}{6.2832}}
\put(3102,297){\arc{210}{0}{1.5708}}
\path(2712,297)(2712,402)
\path(2817,507)(3102,507)
\path(3207,402)(3207,297)
\path(3102,192)(2817,192)
\put(2847,282){\makebox(0,0)[lb]{\smash{{{\SetFigFont{9}{10.8}{\rmdefault}{\mddefault}{\updefault}$O_n$}}}}}
\put(1917,297){\arc{210}{1.5708}{3.1416}}
\put(1917,402){\arc{210}{3.1416}{4.7124}}
\put(2202,402){\arc{210}{4.7124}{6.2832}}
\put(2202,297){\arc{210}{0}{1.5708}}
\path(1812,297)(1812,402)
\path(1917,507)(2202,507)
\path(2307,402)(2307,297)
\path(2202,192)(1917,192)
\put(1947,282){\makebox(0,0)[lb]{\smash{{{\SetFigFont{9}{10.8}{\rmdefault}{\mddefault}{\updefault}$O_1$}}}}}
\put(2397,282){\makebox(0,0)[lb]{\smash{{{\SetFigFont{9}{10.8}{\rmdefault}{\mddefault}{\updefault}.....}}}}}
\put(2097,2457){\arc{210}{1.5708}{3.1416}}
\put(2097,2562){\arc{210}{3.1416}{4.7124}}
\put(2337,2562){\arc{210}{4.7124}{6.2832}}
\put(2337,2457){\arc{210}{0}{1.5708}}
\path(1992,2457)(1992,2562)
\path(2097,2667)(2337,2667)
\path(2442,2562)(2442,2457)
\path(2337,2352)(2097,2352)
\put(2127,2442){\makebox(0,0)[lb]{\smash{{{\SetFigFont{9}{10.8}{\rmdefault}{\mddefault}{\updefault}$G$}}}}}
\path(4737,2526)(6312,2526)
\put(252,2547){\arc{210}{1.5708}{3.1416}}
\put(252,2652){\arc{210}{3.1416}{4.7124}}
\put(492,2652){\arc{210}{4.7124}{6.2832}}
\put(492,2547){\arc{210}{0}{1.5708}}
\path(147,2547)(147,2652)
\path(252,2757)(492,2757)
\path(597,2652)(597,2547)
\path(492,2442)(252,2442)
\put(1017,1287){\arc{210}{1.5708}{3.1416}}
\put(1017,1392){\arc{210}{3.1416}{4.7124}}
\put(1302,1392){\arc{210}{4.7124}{6.2832}}
\put(1302,1287){\arc{210}{0}{1.5708}}
\path(912,1287)(912,1392)
\path(1017,1497)(1302,1497)
\path(1407,1392)(1407,1287)
\path(1302,1182)(1017,1182)
\put(4827,2352){\makebox(0,0)[lb]{\smash{{{\SetFigFont{7}{8.4}{\rmdefault}{\mddefault}{\itdefault}E-method declaration}}}}}
\put(1047,1272){\makebox(0,0)[lb]{\smash{{{\SetFigFont{9}{10.8}{\rmdefault}{\mddefault}{\updefault}$E_n$}}}}}
\put(822,2667){\makebox(0,0)[lb]{\smash{{{\SetFigFont{7}{8.4}{\rmdefault}{\mddefault}{\itdefault}DeleteEdge}}}}}
\path(3972,1272)(5457,1272)
\put(597,1272){\makebox(0,0)[lb]{\smash{{{\SetFigFont{9}{10.8}{\rmdefault}{\mddefault}{\updefault}.....}}}}}
\put(2847,2037){\makebox(0,0)[lb]{\smash{{{\SetFigFont{7}{8.4}{\rmdefault}{\mddefault}{\itdefault}E-method invocation}}}}}
\put(4107,1047){\makebox(0,0)[lb]{\smash{{{\SetFigFont{7}{8.4}{\rmdefault}{\mddefault}{\itdefault}E-method definition}}}}}
\path(2037,507)(2168,2347)
\blacken\path(2189.402,2225.173)(2168.000,2347.000)(2129.554,2229.433)(2189.402,2225.173)
\path(5637,372)(7032,372)
\put(5727,147){\makebox(0,0)[lb]{\smash{{{\SetFigFont{7}{8.4}{\rmdefault}{\mddefault}{\itdefault}E-method definition}}}}}
\path(2937,507)(2333,2347)
\blacken\path(2398.930,2242.342)(2333.000,2347.000)(2341.923,2223.629)(2398.930,2242.342)
\dashline{60.000}(4737,2487)(2442,2487)
\blacken\path(2562.000,2517.000)(2442.000,2487.000)(2562.000,2457.000)(2562.000,2517.000)
\dashline{60.000}(507,2127)(2802,2127)(2802,1722)
	(507,1722)(507,2127)
\thicklines
\path(2532,2037)(2622,1767)
\blacken\thinlines
\path(2555.592,1871.355)(2622.000,1767.000)(2612.513,1890.329)(2555.592,1871.355)
\put(237,2532){\makebox(0,0)[lb]{\smash{{{\SetFigFont{9}{10.8}{\rmdefault}{\mddefault}{\updefault}$X$}}}}}
\put(117,1287){\arc{210}{1.5708}{3.1416}}
\put(117,1392){\arc{210}{3.1416}{4.7124}}
\put(402,1392){\arc{210}{4.7124}{6.2832}}
\put(402,1287){\arc{210}{0}{1.5708}}
\path(12,1287)(12,1392)
\path(117,1497)(402,1497)
\path(507,1392)(507,1287)
\path(402,1182)(117,1182)
\put(147,1272){\makebox(0,0)[lb]{\smash{{{\SetFigFont{9}{10.8}{\rmdefault}{\mddefault}{\updefault}$E_1$}}}}}
\thicklines
\path(2082,2037)(2050,1765)
\blacken\thinlines
\path(2034.226,1887.683)(2050.000,1765.000)(2093.815,1880.673)(2034.226,1887.683)
\dashline{60.000}(5637,327)(3207,327)
\blacken\path(3327.000,357.000)(3207.000,327.000)(3327.000,297.000)(3327.000,357.000)
\path(282,1497)(2003,2417)
\blacken\path(1911.315,2333.970)(2003.000,2417.000)(1883.029,2386.884)(1911.315,2333.970)
\path(1227,1497)(2038,2377)
\blacken\path(1978.738,2268.428)(2038.000,2377.000)(1934.617,2309.089)(1978.738,2268.428)
\dashline{60.000}(3972,1317)(1407,1317)
\blacken\path(1527.000,1347.000)(1407.000,1317.000)(1527.000,1287.000)(1527.000,1347.000)
\thicklines
\path(597,2577)(1992,2577)
\blacken\thinlines
\path(1872.000,2547.000)(1992.000,2577.000)(1872.000,2607.000)(1872.000,2547.000)
\thicklines
\path(1632,2037)(1407,1812)
\blacken\thinlines
\path(1470.640,1918.066)(1407.000,1812.000)(1513.066,1875.640)(1470.640,1918.066)
\thicklines
\path(1047,2037)(687,1857)
\blacken\thinlines
\path(780.915,1937.498)(687.000,1857.000)(807.748,1883.833)(780.915,1937.498)
\path(5502,2217)(5367,1992)(5637,1992)(5502,2217)
\path(3972,1587)(5457,1587)(5457,912)
	(3972,912)(3972,1587)
\path(5637,687)(7032,687)(7032,12)
	(5637,12)(5637,687)
\put(5817,462){\makebox(0,0)[lb]{\smash{{{\SetFigFont{9}{10.8}{\rmdefault}{\bfdefault}{\updefault}Orientation}}}}}
\put(5277,2605){\makebox(0,0)[lb]{\smash{{{\SetFigFont{9}{10.8}{\rmdefault}{\bfdefault}{\updefault}Graph}}}}}
\path(4737,2802)(6312,2802)(6312,2217)
	(4737,2217)(4737,2802)
\path(4692,1587)(4692,1767)(6222,1767)(6222,687)
\path(5502,1767)(5502,1992)
\put(4242,1362){\makebox(0,0)[lb]{\smash{{{\SetFigFont{9}{10.8}{\rmdefault}{\bfdefault}{\updefault}Embedding}}}}}
\put(5465,2029){\makebox(0,0)[lb]{\smash{{{\SetFigFont{5}{6.0}{\rmdefault}{\bfdefault}{\updefault}E}}}}}
\end{picture}
}}}

    \caption{In this mixed (classes-objects) diagram, dashed lines
    represent instance relationships ($E_i$ are embeddings
    and $O_j$ are orientations), thick arrows represent method
    (or E-method) invocations, plain lines represent extension
    relationships. When a method is invoked on a {\cn
    Graph} $G$ (for example \emph{DeleteEdge}), the E-method mechanism
    can be used to notify the event to the current extender-objects. }

    \label{fi-extender-simple-example} 

    \end{figure}
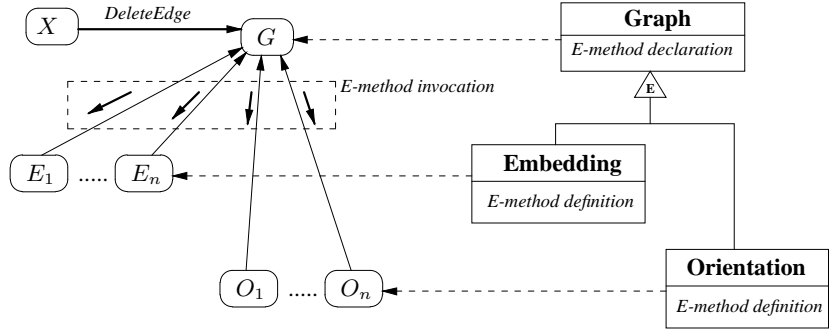
\fi

Fig.~\ref{fi-emethods-interaction} shows the
evolution of a system in which a graph $G$ (a
support-object) and some embedding and orientation
$E_1,\dots,E_n,O_1,\dots,O_m$ for $G$ (extension-objects) interact. The
vertical direction represents the time. Each vertical line represents the
evolution of an object. A vertical line becomes thick when a method is
executed on the corresponding object. The E-method
Post\_AddVertex($v$) is invoked from a method AddVertex(~) of
$G$. This triggers the execution on each embedding and each orientation
of the behavior of Post\_AddVertex(~) associated to such
objects. Observe that distinct extension-objects may have a distinct
behavior depending on the extenders they belong to.  So, in the
example, orientations can have a behavior that is different from the behavior of
embeddings.

\ifmac
\else
    \begin{figure}
    \centering{\centering{\setlength{\unitlength}{0.00078740in}
\begingroup\makeatletter\ifx\SetFigFont\undefined%
\gdef\SetFigFont#1#2#3#4#5{%
  \reset@font\fontsize{#1}{#2pt}%
  \fontfamily{#3}\fontseries{#4}\fontshape{#5}%
  \selectfont}%
\fi\endgroup%
{\renewcommand{\dashlinestretch}{30}
\begin{picture}(5414,2322)(0,-10)
\dottedline{45}(3825,1497)(4185,1497)
\dottedline{45}(3825,867)(4185,867)
\dottedline{45}(3870,147)(4230,147)
\path(4545,192)(4545,12)
\path(4500,372)(4590,372)(4590,192)
	(4500,192)(4500,372)
\path(4545,1767)(4545,372)
\path(3825,1992)(3824,1991)(3821,1987)
	(3814,1977)(3804,1964)(3793,1950)
	(3782,1938)(3772,1930)(3763,1924)
	(3753,1921)(3745,1919)(3739,1919)
	(3734,1919)(3727,1920)(3715,1921)
	(3697,1922)(3669,1924)(3655,1925)
	(3639,1926)(3621,1927)(3601,1929)
	(3580,1931)(3557,1933)(3533,1935)
	(3508,1938)(3482,1940)(3455,1943)
	(3429,1945)(3403,1947)(3377,1950)
	(3353,1952)(3329,1954)(3307,1955)
	(3286,1956)(3267,1957)(3249,1958)
	(3233,1958)(3209,1958)(3187,1956)
	(3168,1954)(3151,1952)(3135,1949)
	(3121,1946)(3106,1942)(3093,1939)
	(3080,1935)(3068,1932)(3056,1928)
	(3046,1924)(3035,1917)(3027,1908)
	(3022,1896)(3019,1883)(3017,1871)
	(3015,1862)(3015,1858)(3015,1857)
\path(3825,1992)(3826,1991)(3829,1987)
	(3836,1977)(3846,1964)(3857,1950)
	(3868,1938)(3878,1930)(3887,1924)
	(3897,1921)(3905,1919)(3911,1919)
	(3916,1919)(3923,1920)(3935,1921)
	(3953,1922)(3981,1924)(3995,1925)
	(4011,1926)(4029,1927)(4049,1929)
	(4070,1931)(4093,1933)(4117,1935)
	(4142,1938)(4168,1940)(4195,1943)
	(4221,1945)(4247,1947)(4273,1950)
	(4297,1952)(4321,1954)(4343,1955)
	(4364,1956)(4383,1957)(4401,1958)
	(4417,1958)(4441,1958)(4463,1956)
	(4482,1954)(4499,1952)(4515,1949)
	(4529,1946)(4544,1942)(4557,1939)
	(4570,1935)(4582,1932)(4594,1928)
	(4604,1924)(4615,1917)(4623,1908)
	(4628,1896)(4631,1883)(4633,1871)
	(4635,1862)(4635,1858)(4635,1857)
\path(1485,1227)(3105,1227)
\blacken\path(2985.000,1197.000)(3105.000,1227.000)(2985.000,1257.000)(2985.000,1197.000)
\path(1485,1137)(3465,1137)
\blacken\path(3345.000,1107.000)(3465.000,1137.000)(3345.000,1167.000)(3345.000,1107.000)
\path(1440,1767)(1440,1497)
\path(720,1497)(1395,1497)
\blacken\path(1275.000,1467.000)(1395.000,1497.000)(1275.000,1527.000)(1275.000,1467.000)
\path(3150,1767)(3150,1227)
\path(3105,1227)(3195,1227)(3195,1137)
	(3105,1137)(3105,1227)
\path(3150,1137)(3150,12)
\path(1395,12)(1395,1497)(1485,1497)(1485,12)
\path(3510,1767)(3510,1137)
\path(3510,957)(3510,12)
\path(3465,1137)(3555,1137)(3555,957)
	(3465,957)(3465,1137)
\path(1485,372)(4500,372)
\blacken\path(4380.000,342.000)(4500.000,372.000)(4380.000,402.000)(4380.000,342.000)
\dottedline{45}(2160,507)(2160,1047)
\path(180,1137)(180,462)(135,462)
	(225,327)(315,462)(270,462)(270,1137)
\put(1305,1902){\makebox(0,0)[lb]{\smash{{{\SetFigFont{11}{13.2}{\rmdefault}{\mddefault}{\itdefault}$G$}}}}}
\put(405,1587){\makebox(0,0)[lb]{\smash{{{\SetFigFont{9}{10.8}{\familydefault}{\mddefault}{\updefault}AddVertex($v$)}}}}}
\put(3015,2127){\makebox(0,0)[lb]{\smash{{{\SetFigFont{11}{13.2}{\rmdefault}{\mddefault}{\itdefault}$E_1,\dots,E_n,O_1,\dots,O_m$}}}}}
\put(1620,1317){\makebox(0,0)[lb]{\smash{{{\SetFigFont{9}{10.8}{\familydefault}{\mddefault}{\updefault}Post\_AddVertex($v$)}}}}}
\put(0,732){\makebox(0,0)[lb]{\smash{{{\SetFigFont{14}{16.8}{\rmdefault}{\mddefault}{\itdefault}t}}}}}
\end{picture}
}}}

    \caption{Interaction between a support-object $G$ and its
    extension-objects $E_1,\dots,E_n,O_1,\dots,O_m$.}

    \label{fi-emethods-interaction} 

    \end{figure}
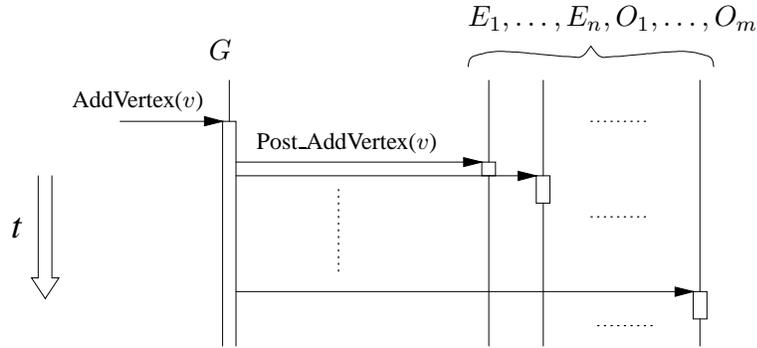
\fi

The E-methods are most likely to be used in methods that change the
state of the support-object. In fact, such changes can lead to
inconsistencies in the current extension-objects. E-methods can
update the extension-object or inhibith the change in the support,
depending on design choices.

\note{Maurizio, guarda se queste modifiche vanno bene}
For each method that performs modifications on the state of the support-object
the designer can provide an E-method that is
called as soon as the method is called.  The definitions of these E-methods
are supposed to inhibit the modification (i.e. throwing an exception)
if the extender constraints are violated. Two more E-methods can be
provided in order to allow each extension-object to update its state
on legal modifications of the graph. These are called before and after
the modification of the support-object.

Extenders are especially suited to support the extension subclassing
where the multiple decoration problem arises (see
Section~\ref{se-multiple-decoration}).  They are useful to
dynamically change behavioral and/or structural aspects of an object
during its life-time. Further, an extender can be developed even much time
after the development of its support-class. This allows to add
capabilities to the system without changing any other part of it. The
new extender will coexist with other extenders already present in the
system, and will work in conjunction with them.

\subsection{Classers}

The {\em classer} is the second new concept. A classer is a
constrained extender for which just one instance is allowed for a
given support-object\footnote{It is important not to misidentify the
constraint for the classer with the \emph{singleton}
pattern~\cite{IB-D963068}. More than one instance is admitted for
classers overall, but only one for each support-object.}. In spite of
the simplicity of the definition, classers are as important as
extenders and give much more expressiveness to the paradigm.

Consider the case when new information, and operations related to
it, have to be added to an object and such information can be univocally
determined by its state by means of a functional dependency (for
example the set of connected components, see
Section~\ref{se-multiple-decoration}).  Also, consider the need of
attaching the new information dynamically. In this case
a dynamic classification system would be needed allowing an object to
change its class during its life-time. Even if extenders seem to be a
good choice to support dynamic classification issues, their capability
of having more than one instance, for each support-object, yields
several disadvantages.

The main problem is that, under the above conditions, all the
extension-objects of a given support-object have the same state,
because the information that they contain functionally depends on the
state of the support-object. Single extension-objects cannot change
independently because of their peculiar semantic. So, having more than
one instance is not useful, inefficient, and conceptually misleading.

Then, we can state that the new information and
operations are conceptually bound to the support-object.
Nevertheless, the programmer has to refer explicitly to the
extension-objects using variables or pointers, as he/she refers to any
other object. Also, when classification hierarchies are non-trivial
the programmer has to deal with much more objects than he/she needs.

Classers are extenders that can have only one instance for each
support-object. Hence, the underlying extension-object can be accessed through
the support-object itself without maintaining an explicit reference to it.
Also, using classers avoids the risk of inefficiency due to multiple
instantiations.

Classers are especially suited to support the restriction subclassing
and the extension subclassing in the case of information univocally
determined by the state of the support-object (see
Section~\ref{se-multiple-decoration}). We would like to point out that
using classers in such situations avoids the crossed
classification problem and the promotion efficiency problem.


\section{Using the ECO Paradigm}\label{se-usage}

In this section we present some techniques especially suited to build
large systems using the ECO paradigm. We provide some examples that
show how ECO addresses the problems mentioned in
Section~\ref{classification-problems} while respecting the
open-closed principle and keeping extendibility and flexibility.

\subsection{Using Classers to Add Structure}
\label{se-add-structure}

A library for combinatorial mathematics cannot handle all the needs that the
user could ask. So, it has to provide powerful and flexible ways to extend
its capabilities without modifying the library itself.

Many features that a non naive user can ask require addictional
structures to be maintained with a graph and updated when the graph
changes (for example, connected components set, block cut-vertex tree,
etc.). Such kind of structures are not basic features (there are many
applications that do not require them) and could be an unnecessary
burden if they are not really needed. For this reason, we prefer not
to insert them into the main graph class.

Classers are particularly useful to address such problem. In fact,
using classers yields the following advantages:

\begin{itemize}

  \item the new information can be attached when needed, so, the user
  is not compelled to choose a cumbersome implementation at the
  instantiation time;

  \item the E-methods mechanism provides a clean way to dynamically
  update the structure and permits to place the updating
  code in the classer definition;

  \item the introduction of new classers does not require changes to
  old code and the new and the old classers can be used simultaneously
  without any sort of limitation, thus, obtaining good extendibility and
  flexibility.

\end{itemize}

Following the presented approach, classers turn out to be the natural
containers for dynamic algorithms (for a similar, but ad hoc, approach
see~\cite{Italiano98}).

\subsection{Hierarchies of Classers and Extenders}
\label{se-chaining}

    \ifmac
     \else
         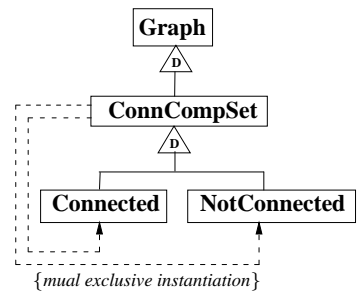
\begin{wrapfigure}{R}{6cm}
         \centering{ \centering{\setlength{\unitlength}{0.00048119in}
\begingroup\makeatletter\ifx\SetFigFont\undefined%
\gdef\SetFigFont#1#2#3#4#5{%
  \reset@font\fontsize{#1}{#2pt}%
  \fontfamily{#3}\fontseries{#4}\fontshape{#5}%
  \selectfont}%
\fi\endgroup%
{\renewcommand{\dashlinestretch}{30}
\begin{picture}(3669,2997)(0,-10)
\dashline{60.000}(822,1935)(12,1935)(12,225)
	(2622,225)(2622,675)
\blacken\path(2652.000,555.000)(2622.000,675.000)(2592.000,555.000)(2652.000,555.000)
\dashline{60.000}(822,1800)(147,1800)(147,360)
	(912,360)(912,675)
\blacken\path(942.000,555.000)(912.000,675.000)(882.000,555.000)(942.000,555.000)
\put(192,0){\makebox(0,0)[lb]{\smash{{{\SetFigFont{7}{8.4}{\rmdefault}{\mddefault}{\itdefault}$\{$mual exclusive instantiation$\}$}}}}}
\path(1731,2579)(1560,2295)(1902,2295)(1731,2579)
\put(1684,2342){\makebox(0,0)[lb]{\smash{{{\SetFigFont{5}{6.0}{\rmdefault}{\bfdefault}{\updefault}D}}}}}
\path(1713,1710)(1542,1426)(1884,1426)(1713,1710)
\put(1666,1473){\makebox(0,0)[lb]{\smash{{{\SetFigFont{5}{6.0}{\rmdefault}{\bfdefault}{\updefault}D}}}}}
\path(282,1024)(1632,1024)(1632,675)
	(282,675)(282,1024)
\put(398,802){\makebox(0,0)[lb]{\smash{{{\SetFigFont{9}{10.8}{\rmdefault}{\bfdefault}{\updefault}Connected}}}}}
\path(1857,1024)(3657,1024)(3657,675)
	(1857,675)(1857,1024)
\put(1332,2686){\makebox(0,0)[lb]{\smash{{{\SetFigFont{9}{10.8}{\rmdefault}{\bfdefault}{\updefault}Graph}}}}}
\path(2674,1218)(2674,1024)
\path(912,1035)(919,1215)
\path(1722,1232)(1722,1440)
\path(822,2034)(2712,2034)(2712,1710)
	(822,1710)(822,2034)
\put(1002,1800){\makebox(0,0)[lb]{\smash{{{\SetFigFont{9}{10.8}{\rmdefault}{\bfdefault}{\updefault}ConnCompSet}}}}}
\path(1722,2295)(1718,2034)
\path(910,1218)(2674,1218)
\put(1991,796){\makebox(0,0)[lb]{\smash{{{\SetFigFont{9}{10.8}{\rmdefault}{\bfdefault}{\updefault}NotConnected}}}}}
\path(1275,2970)(2127,2970)(2127,2571)
	(1275,2571)(1275,2970)
\end{picture}
}}}

         \caption{ {\cn ConnCompSet} is a classer that dynamically
         manage the set of connected components. {\cn Connected} and
         {\cn NotConnected} classers represent the status of the
         graph. They act like an ``hook'' for attaching new classers
         or extenders that deals only with graphs for which the
         relative property holds.}

         \label{fi-prosys-simple-example}
         \end{wrapfigure}
     \fi

In Section~\ref{sec-extender}, we already provided an example of using
extenders to model the concepts of orientation and embedding
(permitting more instances at once of both), and in
Section~\ref{se-add-structure} we described how to use classers.  Now,
we describe how each extender and classer can be a support-class, as
well, accepting its own extenders and classers, and thus, allowing even
more expressiveness.

An extender can be used to associate new structures to an embedding in
order, for example, to represent the orthogonal shape of the
edges\footnote{In the Graph Drawing area, the sequence of left/right
bends of an orthogonal drawing. The shape of a graph plays an
important role in algorithms like GIOTTO~\cite{tamassia:88}.}, and an
orientations can be equipped with costs and capacities to became a
flow network in conjunction with its support graph.

Using this technique the extender-objects can make up a tree.  When a
support-object changes its state, the consistancy in the lower levels
of the extension-objects tree have to be maintained.  E-methods can be
easily used to address this problem.

The depth of such tree is usually not greater then three or four
levels, and constant in any case (it is statically determined at
compile time). On the other hand, there is no constraint to the degree
of its nodes (i.e.~the number of extension-objects). However, note
that the number of extension-objects for a given support-object are
very often constant for a given algorithm and, more precisely, it does
not depend on the input size\footnote{Using a set of extension-objects whose
size grows with the input size is almost always a misuse. In fact,
using extenders implies dynamical update through E-methods, which, in
this case, takes linear time. Note that if we really need to update
such a set of objects dynamically, extenders largely simplify the work
and do not increase the computational complexity.}.  Due to the above
considerations, the paradigm allows the update to be performed in
constant time for a given algorithm.

The classers can also be used to represent restriction subclassing. In
fact, we can associate a classer to a property and the presence of a
classer instance means that the given property holds for the graph,
hence, the graph belong to a specific mathematical class. In
Fig.~\ref{fi-prosys-simple-example} we shows a design schema that
permits to dynamically attach the connected component set structure,
and then (only if such feature is present) to classify the graph as
{\cn Connected} or {\cn NotConnected}.  The instantiation of such
classers can be ascribed to the {\cn ConnCompSet} classer itself. In
this way, we can provide a full-fledged module to handle
connectivity with automatic dynamic classification\footnote{At least
two design choices are possible to deal with operation that disconnect
the graph: an exeption can be thrown or the graph can be dynamically
reclassified as {\cn NotConnected}.}, that also permits further
restriction subclassing (attaching classers to {\cn Connected} or {\cn
NotConnected}), or extenders that can futher exploit the concept of
connected component (attaching extenders to {\cn ConnCompSet}).

\section{Supporting the ECO Paradigm with a Pre-Compiler}
\label{se-supporting}

The introduction of a new paradigm is of little help if not supported
by suitable programming tools. Among several possible alternatives, we
have chosen to support the ECO paradigm with a pre-compiler. Also, we
introduced an extension of the C++ language which we call ECO C++.  The
pre-compiler generates code where new constructs are
replaced by standard C++ code that emulates their semantic.
Fig.~\ref{fi-how-precomp-acts} shows the entire compilation process.

     \ifmac
     \else
         \begin{figure}
         \centering 
         \centering{\setlength{\unitlength}{0.00050000in}
\begingroup\makeatletter\ifx\SetFigFont\undefined%
\gdef\SetFigFont#1#2#3#4#5{%
  \reset@font\fontsize{#1}{#2pt}%
  \fontfamily{#3}\fontseries{#4}\fontshape{#5}%
  \selectfont}%
\fi\endgroup%
{\renewcommand{\dashlinestretch}{30}
\begin{picture}(8767,2361)(0,-10)
\put(3750,1479){\makebox(0,0)[lb]{\smash{{{\SetFigFont{7}{8.4}{\rmdefault}{\mddefault}{\updefault}ECO C++}}}}}
\put(3750,1254){\makebox(0,0)[lb]{\smash{{{\SetFigFont{7}{8.4}{\rmdefault}{\mddefault}{\updefault}precompiler}}}}}
\put(2550,504){\makebox(0,0)[lb]{\smash{{{\SetFigFont{7}{8.4}{\rmdefault}{\mddefault}{\updefault}ECO C++ source}}}}}
\put(2700,279){\makebox(0,0)[lb]{\smash{{{\SetFigFont{7}{8.4}{\rmdefault}{\mddefault}{\updefault}preprocessed}}}}}
\put(0,1104){\makebox(0,0)[lb]{\smash{{{\SetFigFont{7}{8.4}{\familydefault}{\mddefault}{\updefault}ECO C++}}}}}
\put(75,879){\makebox(0,0)[lb]{\smash{{{\SetFigFont{7}{8.4}{\familydefault}{\mddefault}{\updefault}source}}}}}
\path(600,1404)(1200,1404)
\blacken\path(1080.000,1374.000)(1200.000,1404.000)(1080.000,1434.000)(1080.000,1374.000)
\path(3000,1404)(3525,1404)
\blacken\path(3405.000,1374.000)(3525.000,1404.000)(3405.000,1434.000)(3405.000,1374.000)
\path(5250,1404)(5775,1404)
\blacken\path(5655.000,1374.000)(5775.000,1404.000)(5655.000,1434.000)(5655.000,1374.000)
\path(3300,729)(3300,1179)
\blacken\path(3330.000,1059.000)(3300.000,1179.000)(3270.000,1059.000)(3330.000,1059.000)
\path(1200,1704)(3000,1704)(3000,1104)
	(1200,1104)(1200,1704)
\path(3600,1704)(5175,1704)(5175,1104)
	(3600,1104)(3600,1704)
\path(3525,1779)(5250,1779)(5250,1029)
	(3525,1029)(3525,1779)
\path(5775,1704)(7650,1704)(7650,1104)
	(5775,1104)(5775,1704)
\path(7650,1404)(8025,1404)
\blacken\path(7905.000,1374.000)(8025.000,1404.000)(7905.000,1434.000)(7905.000,1374.000)
\path(5475,729)(5475,1179)
\blacken\path(5505.000,1059.000)(5475.000,1179.000)(5445.000,1059.000)(5505.000,1059.000)
\put(2250,2154){\makebox(0,0)[lb]{\smash{{{\SetFigFont{10}{12.0}{\rmdefault}{\bfdefault}{\updefault}ECO C++ compilation process}}}}}
\put(1350,1254){\makebox(0,0)[lb]{\smash{{{\SetFigFont{7}{8.4}{\rmdefault}{\mddefault}{\updefault}preprocessor}}}}}
\put(1350,1479){\makebox(0,0)[lb]{\smash{{{\SetFigFont{7}{8.4}{\rmdefault}{\mddefault}{\updefault}Standard C++}}}}}
\put(5925,1479){\makebox(0,0)[lb]{\smash{{{\SetFigFont{7}{8.4}{\rmdefault}{\mddefault}{\updefault}Standard C++}}}}}
\put(5925,1254){\makebox(0,0)[lb]{\smash{{{\SetFigFont{7}{8.4}{\rmdefault}{\mddefault}{\updefault}compiler}}}}}
\put(7875,1104){\makebox(0,0)[lb]{\smash{{{\SetFigFont{7}{8.4}{\rmdefault}{\mddefault}{\updefault}object code}}}}}
\put(5025,504){\makebox(0,0)[lb]{\smash{{{\SetFigFont{7}{8.4}{\rmdefault}{\mddefault}{\updefault}C++ source}}}}}
\put(4920,54){\makebox(0,0)[lb]{\smash{{{\SetFigFont{7}{8.4}{\rmdefault}{\mddefault}{\updefault}preprocessing}}}}}
\put(4890,294){\makebox(0,0)[lb]{\smash{{{\SetFigFont{7}{8.4}{\rmdefault}{\mddefault}{\updefault}It does not need}}}}}
\end{picture}
}}

         \caption{The ECO C++ compilation process. A pre-compilation phase
         has been inserted after the preprocessing phase and before the
         usual compilation phase.}

         \label{fi-how-precomp-acts}
         \end{figure}
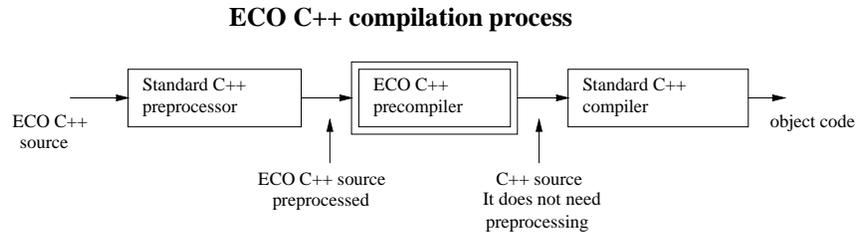
     \fi

\note{forse possiamo togliere questo}
This yields a number of advantages. The C++ language is a well know
language with many libraries and tools already available and widely
used. A pre-compiler permits to use the new paradigm without limiting
the usage of such existing tools. Further, it does not require the
programmer to learn a completely new language, and because of the
existence of good, portable and freely available compilers, the system
works on a wide range of platforms.

\subsection{The ECO C++ language}

ECO C++ supports the concept of extenders and E-methods by means of new
syntactic primitives.  The programmer can declare a support-class and
an extender by using a suitable syntax as in the example shown in
Fig.~\ref{fi-ecocpp-extender-decl}. The keyword {\kw extensible}
denotes a declaration of a support-class in which some E-methods can
be declared.  E-methods are denoted by the keyword {\kw
extend}. The behavior for such E-methods may be defined for each extender
(the default behavior is ``do nothing'').  The E-methods invocation is
performed by using the keyword {\kw call\_e\_method(~)} as shown in
the same figure.  When the control flow reaches such keyword the
execution is dispatched to the behaviors specified in each current
extension-object.

The keyword {\kw extend} denotes an extender for a specified class,
when placed in the head of a class declaration. Each constructor for
the extender must have, as first parameter, a reference or a
pointer to the support-class.  The actual parameter will be the
support-object of the extender-object that is being creating (see
Fig.~\ref{fi-ecocpp-extender-decl}). The programmer can ignore it,
because the pre-compiler takes the responsibility to update the
internal structures. The behavior for the E-methods can be defined by
the programmer as he/she does for usual methods.

\begin{figure}

\begin{tabular}{c|c}
\begin{minipage}[b]{7cm}
\input support-class-example.tex
\end{minipage}
\ \ \ \ \ & \ \ \ \ \ 
\begin{minipage}[b]{7cm}
\input extender-class-example.tex
\end{minipage}
\\
\end{tabular}

\caption{An example of declaration of a support-class {\cn Graph} with an 
         extender {\cn Labeling}, written in ECO C++.}

\label{fi-ecocpp-extender-decl}
\end{figure}

The usage of extenders is quite easy. As it can be noticed in
Fig.~\ref{fi-ecocpp-extender-usage}, extenders are instantiated as
usual classes, but for the special meaning of the first parameter in
the constructor. Further, the extension-objects must be destroyed
before the support-object. Note that this automatically happens if the
storage class of the instance is {\kw auto} (i.e.\ the allocation is
performed on the stack) because the destruction order for the objects
declared in a block is the reverse of the declaration order, according
to the standard C++ semantic~\cite{c++:ansi,Stroustrup91}.

\lagrind{extender-usage.tex}{An example of usage of the extender and the
support-class declared in
Fig.~\protect{\ref{fi-ecocpp-extender-usage}}.}{fi-ecocpp-extender-usage}

ECO C++ supports the classer concept by means of the keyword {\kw
dynamic}.  Fig.~\ref{fi-ecocpp-classer-decl} shows a declaration for a
classer. All constructors have to be private because the presence of
the classer states that a given property holds. Here, we suggest to
introduce a class method (declared {\kw static}) that performs the
test and eventually instantiates the classer, if admissible. We call
such class method a pseudo-constructor. The programmer does not need
to maintain references to classer instances. A special syntax makes easy
to access methods of a classer and to test if a classer is
instantiated for a given support-object:
\begin{center}
  \parbox{16cm}{
   \begin{tabbing}
   {\kw $support\_object$.\{$classer$\}.$method(\ldots )$} \= \qquad \=  call 
the method \\
   {\kw $support\_object$.\{$classer$\} } \> \>return true if the classer is 
instantiated.
   \end{tabbing}
  }
\end{center}
Fig.~\ref{fi-ecocpp-classer-usage} shows the usage of the classer
declared in Fig.~\ref{fi-ecocpp-classer-decl}. It shows how to invoke
a pseudo-constructor, how to test if a classer is instantiated, and
how to preform a classer method calling.  

\note{Si potrebbero togliere
le fig.~\ref{fi-ecocpp-classer-usage} e
        \ref{fi-ecocpp-extender-usage}}

\lagrind{classer-decl.tex}{A delcaration of a classer for managing the planarity
property.}{fi-ecocpp-classer-decl}
\lagrind{classer-usage.tex}{An example of usage of the classer declared
in Fig.~\protect{\ref{fi-ecocpp-classer-decl}}.}{fi-ecocpp-classer-usage}

\note{If I change to ``list form'' I have to write more!}
Similar concepts are implemented in other systems,
but they differ from ECO in substantial aspects. The concepts of
\emph{signal} and \emph{slot}, in Qt~\cite{Qt}, are a support for
dynamic updating comparable to the E-methods system, but they were
born in a quite different field (GUI object component) where no needs
for complex classification arise. So, signals and slots alone are not 
useful in solving the exposed classification problems.  On the
other hand, the Java member classes~\cite{Gosling:1996:JLS}, are
oriented to classification problems, but they lack of dynamic updating
capabilities.  In some sense, ECO represents a fusion of these two
attracting approaches that permits completely new developments.

\subsection{The run-time support of ECO C++}
\label{se-implementation}

\note{Togliamo questo paragrafo? pero' e' molto bello}
The run-time support for the ECO C++ language has to maintain low level
structures that represent the relationships 
between each support-object and its current
extension-objects.  The following operations have to be performed
using such structures:

\begin{itemize}
  \item creation of an extension-object,
  \item deletion of an extension-object,
  \item invocation of an E-method.
\end{itemize}

It is easy to implement a support system that is based on linked lists and
performs a creation and a deletion taking $O(1)$ time. 

On the contrary, the invocation of an E-method takes $O(k)$ time,
where $k$ is the number of the current extension-objects, if the
E-method execution takes $O(1)$ for each extension-object, which is
the most common situation. However, $k$ is most of the time only
dependent on the algorithm, as it arises from the analysis that we made in
Section~\ref{se-chaining}.


\section{Conclusions}

Developing libraries of algorithms is a complex task requiring expertise
both in algorithmics and in software engineering.

In this paper we have presented techniques that are mainly related to solving 
graph classification problems in libraries of graph algorithms.
The introduced representation primitives allow the
designer to model dynamically changing properties of graphs with
little effort. They permit to overcome most of the limitations that
come out with standard object-orientation, and to reach many of its
aimed goals, like extensibility (the open-closed principle) and
flexibility (reuse of preexistent modules in different contexts).
Further, the primitives allow to elegantly (in our opinion) express known
concepts like the \emph{observer pattern}~\cite{IB-D963068} and the
\emph{data accessor}~\cite{WeiheOopsla97}.

The pre-compiler for the ECO C++ language permits to easily apply
the new concepts avoiding many of the drawbacks that typically arise
when using new paradigms. The pre-compiler is available on the Web at the address
below:
\begin{center}
\verb|http://www.dia.uniroma3.it/~pizzonia/eco|
\end{center}

The presented ideas have born in the Graph Drawing area and specifically
during the development and usage of the
GDToolkit~\cite{gdtoolkit_man} system and have already been applied within
the same project.




\nocite{BETT94} 

\bibliographystyle{plain}

\ifmac
\input{eco8.bbl}
\else
\bibliography{%
eco,%
geom,%
extra,%
gd97}
\fi

\end{document}